\documentclass[
11pt,%
tightenlines,%
twoside,%
onecolumn,%
nofloats,%
nobibnotes,%
nofootinbib,%
superscriptaddress,%
noshowpacs,%
centertags]%
{revtex4}
\usepackage{ljm}

\begin{document}
\titlerunning{Properties of almost complete local revivals} 
\authorrunning{Ermakov} 

\title{Generalized Almost Complete Revivals in quantum spin chains}

\author{\firstname{I.V.}~\surname{Ermakov}}
\email[E-mail: ]{ermakov1054@yandex.ru}
\affiliation{Skolkovo Institute of Science and Technology, Skolkovo Innovation Center 3, Moscow 143026, Russia.}
\affiliation{Department of Mathematical Methods for Quantum Technologies, Steklov Mathematical Institute of Russian Academy of Sciences, 8 Gubkina St., Moscow 119991, Russia.}
\affiliation{Laboratory for the Physics of Complex Quantum Systems, Moscow Institute of Physics and Technology, Institutsky per. 9, Dolgoprudny, Moscow region, 141700, Russia}



\received{May 03, 2022} 

\begin{abstract} 
The conception of almost complete revivals has been introduced recently. In a quantum many-body system local observable may exhibit an almost complete revival to its maximal value at the predetermined moment of time. In this paper we extend the original procedure such that the revival may be from an arbitrary point on the Bloch sphere to the arbitrary point. Furthermore in the proposed procedure the reviving and collapsing sites are not necessarily the same. We also demonstrate that for spins $S$ higher than $1/2$ almost complete revivals are suppressed as $1/S$. 
\end{abstract}

\maketitle


\section{\label{s1_intro} Introduction }

Local observables of non-integrable quantum many-body systems are expected to quickly reach their equilibrium values. This process of reaching thermal equilibrium is also known as themalization. It's also usually implied that once some observable has reached thermal equilibrium it stays there practically forever exhibiting only meaningless fluctuations upon its equilibrium value. There is however a number of physical mechanisms when thermalization is either slowed down or not present at all. Among such mechanisms are many-body localization \cite{nandkishore2015many}, spin-glasses \cite{rademaker2020slow,stein2013spin}, systems with long-range interactions \cite{gong2013prethermalization,neyenhuis2017observation} or systems with constrains \cite{turner2018weak,bernien2017probing,lin2019exact}. All these mechanisms however are associated with the properties of the system under consideration, an alternative approach is to consider specially designed initial states without putting significant restrictions on the system itself. 

In our recent work \cite{ermakov2021almost} we proposed a mechanism of constructing special initial states such that the selected local observable after relaxation exhibits almost complete revival (ACR) to its initial value at the predetermined moment of time. In a closed many-body system $C=A\cup B$ for a small subsystem $A$ in a pure state the rest of the system $B$ usually serves as a thermal reservoir, which leads to the equilibration of $A$. However the reservoir $B$ can be finely tuned such that at some predetermined revival time $\tau$, subsystem $A$ will be out of equilibrium. So in fact the reservoir can work both ways: thermalize the subsystem or on the contrary push it out of equilibrium. The latter however is exponentially rare event, yet if one can has access all the degrees of freedom of $B$, then the $B$ can be tuned such that this event will take place at any desired time $\tau$.

\begin{figure}
\centering
\includegraphics[width=0.8\columnwidth]{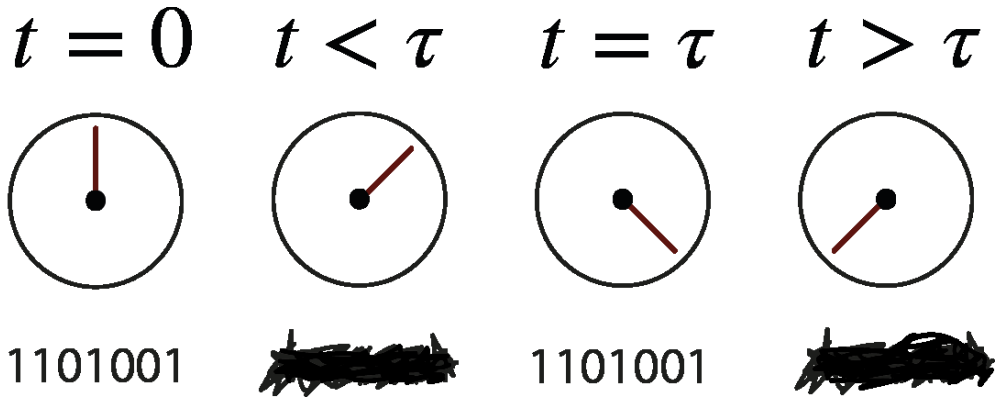}
\caption{Schematic illustration of delayed disclosure of a secret. The information is available only at two moments of time $t=0$ and $t=\tau$. Any attempt to extract the information before the $\tau$ leads to its destruction. 
\label{clocks}}
\end{figure}

In \cite{ermakov2021almost} we also suggested several applications of ACR to benchmarking of quantum similators. In particular, successful preparation of ACR state can be proved with only one local measurement. We had also proposed to utilize ACR for entanglement assisted sensing and delayed disclosure of a secret. The latter is a scheme which allows one to encode a piece of information into the quantum system such that this information will be accessible only in the short vicinity of the revival moment $\tau$ see Figure \ref{clocks}. Any attempt to extract the information earlier will lead to its permanent destruction. In this regard a system of 5-15 qubits prepared in the ACR state represents some sort of ''Quantum time capsule'' which cannot be opened before some specified time. 

In \cite{ermakov2021almost} we considered only revivals of spins polarized along $z$-axis to the same value. In the present paper we provide a general prescription on how to construct ACR from an arbitrary point of a Block Sphere to another arbitrary point. Furthermore the collapsing and reviving spins should not necessarily coincide. In this regard, it is not a revival of some local observable as such, but rather a recovery of purity in local subsystem. 

We also consider spin chains with spin $S>1/2$, and show that in this case ACR are suppressed by factor $1/S$. This result is in agreement with the classical picture where the ACR are impossible due to the fact that classical spin chains are chaotic in general. 

\section{General revival scheme in chains of spin 1/2}

In this section we generalize the prescription offered in \cite{ermakov2021almost} such that the initial and revived values of selected local spin are both arbitrary points on a Bloch Sphere. Furthermore, we show that the revival may occur on a site different from initial one. 

Let us consider a lattice of $L$ interacting spins $1/2$. Each spin is describes by the operator $\{S^\alpha_i\}$ where $i$ is the lattice index and $\alpha=x,y,z$ is the spin projection index. As a collapsing observable we pick the spin $\vec{S}_q$ and $\vec{S}'_p$ as a reviving one. Notice that while values of $q$ and $p$ as well as $\vec{S}_q$ and $\vec{S}'_p$ are different in general, there are no restrictions for them to coincide. 

One-spin Hilbert spaces are defined as $|1_i\rangle$ and $|0_i\rangle$, such that $\langle 1_i|S^z_i| 1_i\rangle=1/2$ and $\langle 0_i|S^z_i| 0_i\rangle=-1/2$. Let us introduce the basis $\mathcal{B}=\{|\varphi_j\rangle\}^{2^L}_{j=1}$, here $|\varphi_j\rangle$ is a many-body basis vector. Let us use the following ordering for the basis $\mathcal{B}$:

\begin{align}\label{basis05}
\mathcal{B}=\{|1_1 \ 1_2  \dots 1_L\rangle, |1_1 \ 1_2  \dots 0_L\rangle, \ \dots \ , |0_1 \ 0_2 \dots 1_L\rangle,|0_1 \ 0_2 \dots 0_L\rangle\}.
\end{align}

Choice of basis determines further construction of ACR states. In our case it is convenient to quantize basis along $z$-axis and order it such as each many-body vector $|\varphi_j\rangle$ corresponds to the base-$2$ form of an integer $j$. 

Let us use the following parametrization of the ACR state: 

\begin{align}\label{ap1}
|\Phi_\text{ACR}(0)\rangle=\sum\limits^{2^{q-1}}_{k=1}\sum\limits^{2^{L-q}}_{n=1}\left(A_{s(k,n)}|\varphi_{s(k,n)}\rangle+\alpha A_{s(k,n)}|\varphi_{s(k,n)+2^{L-q}}\rangle\right),
\end{align}
here $q$ is the collapsing site, $\alpha$ - complex parameter defining the state of the collapsing site on a Bloch Sphere, and function $s(k,n)$ is defined as:

\begin{align}\label{sknf}
s(k,n)=2^{L-q+1}(k-1)+n.
\end{align}

The ansatz (\ref{ap1}) for the initial wavefunction guarantees that it  has a form of tensor product $|\Phi_\text{ACR}(0)\rangle=|l_q\rangle\otimes|\Psi_\text{res}\rangle$, where $|l_q\rangle$ is the wavefunction of the $q$-th site which has the form:

\begin{align}\label{ap2}
|l_q\rangle=\frac{|0_q\rangle+\alpha|1_q\rangle}{\sqrt{1+|\alpha|^2}},
\end{align}

$|\Psi_\text{res}\rangle$ describes the rest of the system, we refer to $|\Psi_\text{res}\rangle$ as to a 'reservoir'. The parameter $\alpha$ is a complex number which determines the position of $q$-th spin on a Bloch sphere as below:

\begin{align}
\langle S^x_q\rangle=\frac{\text{Re}\alpha}{1+|\alpha|^2}, \qquad \langle S^y_q\rangle=\frac{\text{Im}\alpha}{1+|\alpha|^2}, \qquad \langle S^z_q\rangle=-\frac{1}{2}\cdot\frac{1-|\alpha|^2}{1+|\alpha|^2}.\nonumber
\end{align}

\begin{figure}
\centering
\includegraphics[width=0.95\columnwidth]{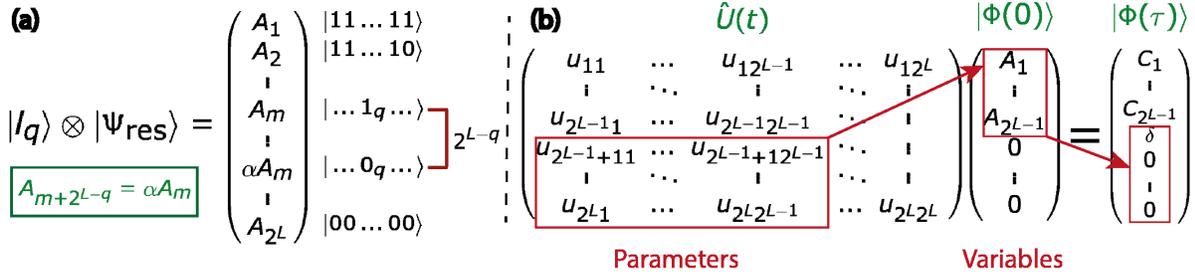}
\caption{(a) Schematic representation of the ansats (\ref{ap1}) in the many-body Hilbert space. (b) Example of revival conditions (\ref{cond2}) in case of $q=p=1$ and $\vec{S}_1=\vec{S}'_1=(0,0,1/2)$. Matrix $\hat{V}$ coincides with the bottom-left submatrix $u$ of size $2^{L-1}$, states (\ref{ap1}) and (\ref{ap3}) have simple structure in the basis (\ref{basis05}). 
\label{scheme}}
\end{figure}

Let us take a closer look at the expression (\ref{ap1}). In general, a wavefunction in a form of tensor product $|l_q\rangle\otimes|\Psi_\text{res}\rangle$ has $2^{L-1}$ independent parameters $\mathcal{A}=\{\{A_{s(k,n)}\}^{2^{L-1}}_{n=1}\}^{2^{q-1}}_{k=1}$. Function (\ref{sknf}) gives us pairs of basis vectors $|\varphi_{s(k,n)}\rangle$ and $|\varphi_{s(k,n)+2^{L-q}}\rangle$ such that these two vectors are identical for all but one $q$-th site. For example if we set $q=L$, then for $k=1$ we will have $|\varphi_{s(1,1)}\rangle=|1_1 \ 1_2  \dots 1_L\rangle$ and $|\varphi_{s(1,1)+1}\rangle=|1_1 \ 1_2  \dots 0_L\rangle$, taking all the $k=\overline{1,2^{q-1}}$ we will end up with $2^{L-1}$ basis vectors which are different only for the $L$-th site. Now to have a $q$-th spin in a pure state (\ref{ap2}) we need to demand that $A_{s(k,n)+2^{L-q}}=\alpha A_{s(k,n)}$. If this condition is satisfied, then initial wavefunction always has a form of tensor product $|\Phi_\text{ACR}(0)\rangle=|l_q\rangle\otimes|\Psi_\text{res}\rangle$ for any set of parameters $\mathcal{A}$. In Figure \ref{scheme} (a) we schematically illustrate the ansatz (\ref{ap1}). 

The set of parameters $\mathcal{A}$ defines the state of the reservoir $|\Psi_\text{res}\rangle$. If all the parameters are chosen randomly then the $q$-th spin will quickly entangle with the reservoir and will remain entangled virtually forever, same applies to all other spins. Our goal is to choose such set $\mathcal{A}$ that at the specified ''revival'' time $\tau$ wavefunction would split into a tensor product again $|\Phi_\text{ACR}(\tau)\rangle=|\bar{l}_p\rangle\otimes|\bar{\Psi}_\text{res}\rangle$. 

Let us demand:

\begin{align}\label{ap3}
|\Phi_\text{ACR}(\tau)\rangle&=e^{-iH\tau}|\Phi_\text{ACR}(0)\rangle=\nonumber\\
&\sum\limits^{2^{p-1}}_{k=1}\sum\limits^{2^{L-p}}_{n=1}\left(C_{\bar{s}(k,n)}|\varphi_{\bar{s}(k,n)}\rangle+\beta C_{\bar{s}(k,n)}|\varphi_{\bar{s}(k,n)+2^{L-p}}\rangle\right),
\end{align}

this wavefunction has a form similar to (\ref{ap1}), with the difference that now it is $p$-th spin in the pure state. The state $|\bar{l}_p\rangle$ is parametrized by the complex number $\beta$ similarly to (\ref{ap2}):

\begin{align}
|l_q\rangle=\frac{|0_q\rangle+\beta|1_q\rangle}{\sqrt{1+|\beta|^2}},\nonumber
\end{align}
we also modified $s(k,n)$ to $\bar{s}(k,n)=2^{L-p+1}(k-1)+n$, to have a revival on $p$-th site. 

To find such set $\mathcal{A}$ that (\ref{ap3}) is satisfied we need to know the full form of the evolution operator at the revival moment $u\equiv e^{-iH\tau}$. Let us take a look at the full form of the condition (\ref{ap3}):

\begin{align}\label{apsys}
&u_{1,1}A_1+\cdots+u_{1,2^L}A_{2^L}=C_1\nonumber\\
&u_{2,1}A_2+\cdots+u_{2,2^L}A_{2^L}=C_2\nonumber\\
&\vdots \\
&u_{2^L,1}A_2+\cdots+u_{2^L,2^L}A_{2^L}=C_{2^L}.\nonumber
\end{align}

By substituting $A_{s(k,n)+2^{L-q}}=\alpha A_{s(k,n)}$ into (\ref{apsys}), for $n=\overline{1,2^{L-q}}$ and $k=\overline{1,2^{q-1}}$, we eliminate $2^{L-1}$ variables from the system (\ref{apsys}). We can also eliminate $2^{L-1}$ equations from (\ref{apsys}) by using the fact that $C_{\bar{s}(k,n)+2^{L-p}}=\alpha C_{\bar{s}(k,n)}$. Thus we obtain a set of conditions: 

\begin{align}\label{cond1}
\hat{V}\mathcal{A}=0,
\end{align}
where the matrix $\hat{V}$ is given by: 

\begin{align}\label{ap4m}
V_{ki}=u_{d[k],\bar{d}[i]}-\beta^{-1} u_{d[k]+2^{L-p},\bar{d}[i]}+\alpha u_{d[k],\bar{d}[i]+2^{L-q}}-\alpha\beta^{-1}u_{d[k]+2^{L-p},\bar{d}[k]+2^{L-q}},
\end{align}

where indexes $k,j=\overline{1,2^{L-1}}$, and sets of indexes $d$ and $\bar{d}$ are ordered sets:

\begin{align}\label{setsindex}
&d=\{\{s(k,n)\}^{2^{q-1}}_{k=1}\}^{2^{L-q}}_{n=1},\nonumber\\
&\bar{d}=\{\{\bar{s}(k,n)\}^{2^{p-1}}_{k=1}\}^{2^{L-p}}_{n=1}.
\end{align}

If matrix (\ref{ap4m}) is degenerate, then (\ref{cond1}) has a solution and therefore (\ref{ap3}) is satisfied exactly. However we argued in \cite{ermakov2021almost} that in case of interacting non-integrable Hamiltonian (\ref{ap4m}) must always be non-degenerate. In this case the only solution of $\mathcal{A}=0$ which is irrelevant. Let us allow one equation from (\ref{cond1}) have non-zero right-hand side

\begin{align}\label{cond2}
\hat{V}\mathcal{A}=(\delta,0,\dots,0)^T,
\end{align}

here $\delta$ is a parameter which will be determined from normalization condition on (\ref{ap2}). 

Let us assume that typical matrix element of $v$ have absolute value $|v_{ki}|\sim 1/\sqrt{2^L}$ and a largely random phase. In this case we can estimate typical values of $\mathcal{A}$ and $\mathcal{C}$ as$|A_0|\sim 1/\sqrt{2^{L-1}}$ and $|C_0|\sim 1/\sqrt{2^{L-1}}$. Now if we substitute these typical values into the left-hand side of the first equation in (\ref{cond2}) we obtain $\delta\sim 1/\sqrt{2^{L-1}}$. This is a key assumption for the existence of ACR, we will discuss it further in the paper. 

Let us formulate the procedure of ACR construction as step by step algorithm. In order to construct ACR one needs to: 

\begin{enumerate}
  \item Pick a collapsing $q$ and reviving $p$ sites.
  \item Choose a position of collapsing $\vec{S}_q$ and reviving $\vec{S}'_p$ spins on a Bloch Sphere. Determine corresponding parameters $\alpha$ and $\beta$. 
  \item Compute sets of indexes $d$ and $\bar{d}$  (\ref{setsindex}).
  \item Compute matrix (\ref{ap4m}).
  \item Solve set of equations (\ref{cond2}). In practice one can set $\delta=1$ to obtain a non-normalized solution. 
  \item When set of parameters $\mathcal{A}$ is determined, construct wavefunction (\ref{ap1}) and normalize it. 
\end{enumerate}

In Figure \ref{scheme} (b) we illustrate a particular example of obtaining conditions (\ref{cond2}). In this example we imply that $p=q$ and that $\vec{S}_p=\vec{S}'_q=(0,0,1/2)$, which corersponds to $\alpha,\beta\rightarrow\infty$ or simply $|l_q\rangle=|1_q\rangle$. In this case matrix (\ref{ap4m}) is simply bottom-left block of size $2^{L-1}$ and wavefunctions (\ref{ap1}) and (\ref{ap3}) are easy to construct. This example corresponds to the one considered in \cite{ermakov2021almost}.

\subsection{Example of ACR}

Let us now construct ACR for the Hamiltonian:

\begin{align}\label{hamhuse}
  H_1= & \sum_{j=1}^L\left(g\sigma^x_j+h\sigma^z_j+J\sigma^z_j\sigma^z_{j+1} \right),
\end{align}
here parameters $(g,h,J)=(0.9045,0.8090,1)$ are used. Periodic boundary conditions $\sigma^\alpha_i=\sigma^\alpha_{i+L}$ are imposed. For the system sizes where exact diaginalization is available, it was tested thoroughly in \cite{kim2013ballistic,kim2014testing} that this Hamiltonian is in a great agreement with Eigenstate Thermalization Hypothesis (ETH) \cite{deutsch1991quantum,srednicki1994chaos}. 

\begin{figure}
\centering
\includegraphics[width=0.8\columnwidth]{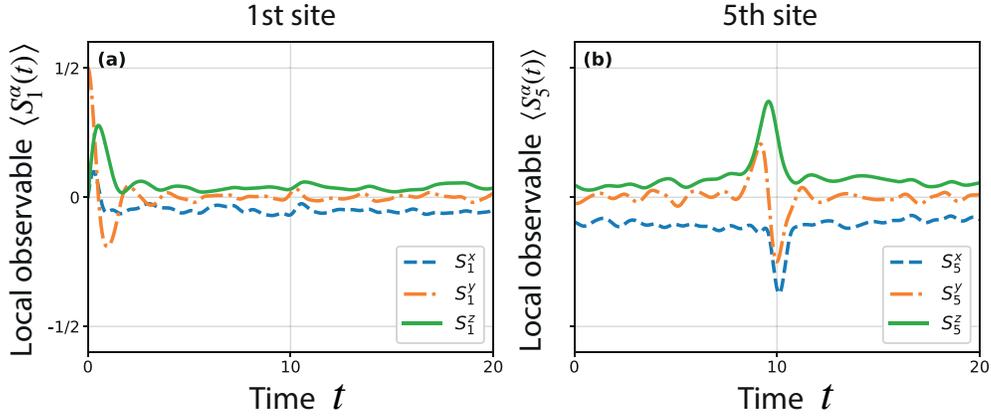}
\caption{Time evolution of local observables $S^\alpha_m$ for the Hamiltonian (\ref{hamhuse}). The dynamics on the 1st site $m=1$ is shown in (a) and of the 5th $m=5$ in (b). Revival time $\tau=10.0$, system size $L=12$, $\alpha=i$, $\beta=-\sqrt{2/9}-1/3i$.
\label{ap_huse_fig}}
\end{figure}

Consider the initial state $|\Phi_\text{ACR}(0)\rangle=|l_1\rangle\otimes|\Psi_\text{res}\rangle$, here $\alpha=i$ therefore $\vec{S}_1=(0,1/2,0)$. Let us set the revival time as $\tau=10$, and pick $p=5$ and $\beta=-\sqrt{2/9}-1/3i$. By solving the system of equations (\ref{cond2}) for the system of $L=12$ spins we find such $|\Phi_\text{ACR}(0)\rangle$ that $\vec{S}_5=(-0.353611,-0.249662,0.249849)$, see Fig. \ref{ap_huse_fig}. The norm of $|\vec{S}_5|^2=0.499797$ is close to $1/2$, therefore the $5$-th spin is almost at the pure state at the revival moment. We showed in \cite{ermakov2021almost} that the discrepancy between perfect revival and ACR vanishes exponentially with the system size. 

\section{Higher spins}~

\begin{figure}
\includegraphics[width=\columnwidth]{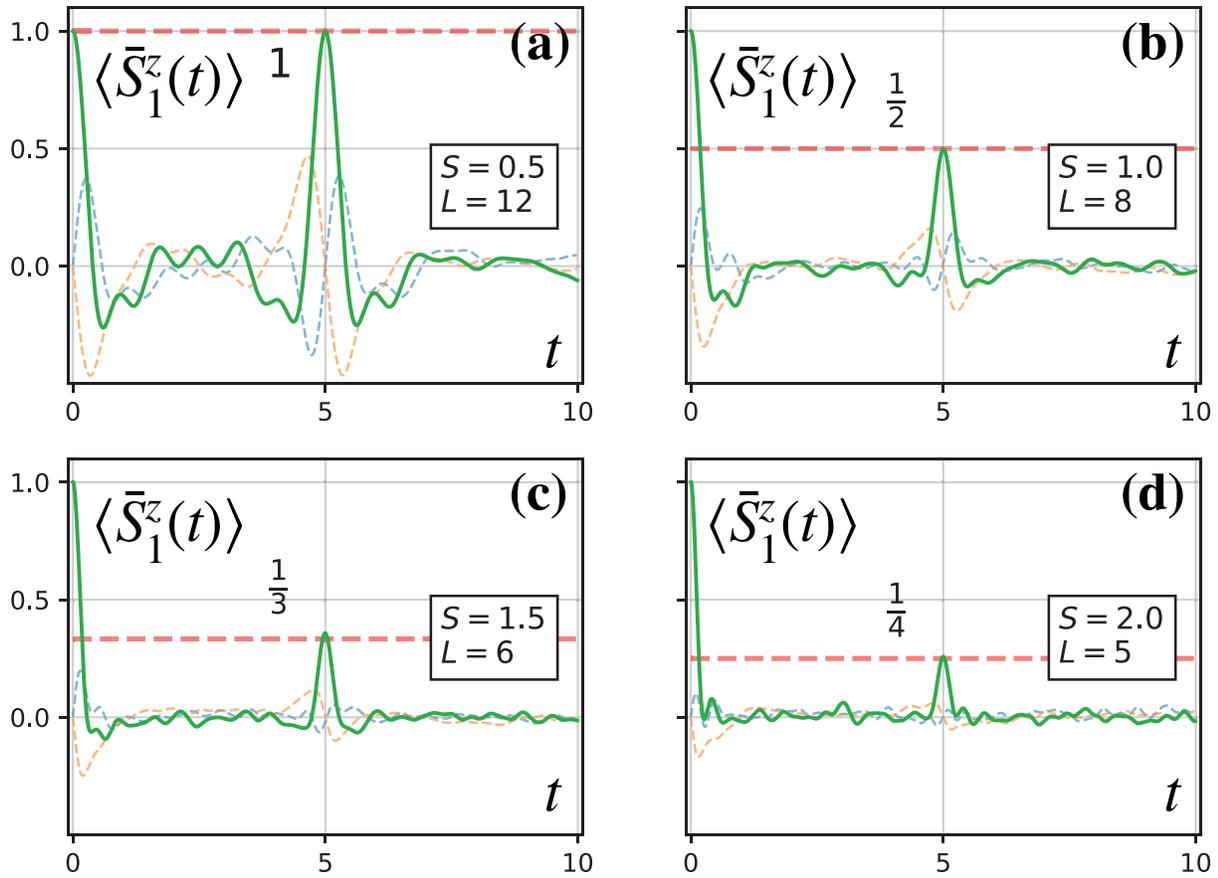}
\caption{Time evolution of the local observables $\langle \bar{S}^\alpha_1\rangle=\langle S^\alpha_1\rangle/S$, for the Hamiltonian (\ref{ap_hams}) for different quantum spins $S=\frac{1}{2},1,\frac{3}{2},2$. Green line corresponds to the $z$ projection, dotted blue and orange lines to $x$ and $y$ correspondingly. Revival time $\tau=5$.
\label{difs}}
\end{figure}

In this section we apply the mechanism of ACR construction for the case of quantum spins $S$ higher than $\frac{1}{2}$. Let us consider the case when collapsing and reviving sites coincide $q=p=1$ and when $|l_1\rangle=|1_1\rangle$. Let us consider the Hamiltonian: 

\begin{align}\label{ap_hams}
  H_2= & \sum_{j=1}^L\left(J_x\, S_j^x S_{j+1}^x +J_y\,S_j^y S_{j+1}^y \right)\nonumber\\
   & +\sum_{j=1}^L\left(h_x\, S_j^x+h_y\, S_j^y \right),
\end{align}

Periodic boundary conditions are imposed and parameters $(J_x,J_y,h_x,h_y)=(-2.0,-4.0,2.2,2.2)$ are used. Since this Hamiltonian acts in XY plane, then the equilibrium values of $\langle S^z_j\rangle=0$. The Hamiltonian (\ref{ap_hams}) is far from integrability, as evidenced by energy-level-spacing statistics \cite{atas2013distribution}.

For spins $S$ it is convenient to order the basis $\mathcal{B}_S=\{|\varphi^S_j\rangle\}^{g^L}_{j=1}$ as $\mathcal{B}_S=\{g^L_g,...,2_g,1_g,0_g\}$ here $g=2S+1$ and $j_g$ is a base-$g$ form of an integer $j$. The initial state $|\Phi_\text{ACR}(0)\rangle$ has the form:
\begin{align}\label{psiinS}
|\Phi_\text{ACR}(0)\rangle=\sum\limits^{g^{L-1}}_{n=1}A_n|\varphi^S_n\rangle.
\end{align}
conditions for observation ACR in the basis $\mathcal{B}_S$ changes to:

\begin{align}\label{sysS}
&\sum\limits^{g^{L-1}}_{n=1}u_{g^{L-1}(g-1)+1,n}A_n=\delta\nonumber \\
&\sum\limits^{g^{L-1}}_{n=1}u_{g^{L-1}(g-1)+2,n}A_n=0\nonumber \\
&\cdots \\
&\sum\limits^{g^{L-1}}_{n=1}u_{g^L,n}A_n=0.\nonumber
\end{align}

The system (\ref{sysS}) has $g^{L-1}$ variables and equations, $\delta\neq 0$ to be determined from normalization conditions. The dimensionality of the Hilbert space is $\mathcal{D}=g^L$, the system (\ref{psiinS}) allows us to set to zero at the moment $\tau$ only $g^{L-1}-1$ of the coefficients. There are also $p=g^{L-1}(g-1)+1$ of non-zero coefficients left in $e^{-iHt}|\Phi_\text{ACR}(0)\rangle$, so $\langle S^z_1(\tau)\rangle$ can not exhibit almost complete revival for spins higher than $S>\frac{1}{2}$. With increasing system size its revival value converges to:

\begin{align}
\langle S^z_1(\tau)\rangle\simeq\frac{1}{2S}.\nonumber
\end{align}

In a Fig. \ref{difs} we plot the time evolution of $\langle \bar{S}^\alpha_1\rangle=\langle S^\alpha_1\rangle/S$ for different values of spin $S$. The $\langle\bar{S}^z_1(\tau)\rangle$ decreases with $S$ which is in agreement with the classical picture in which one can not predict trajectory for arbitrary time if the system is chaotic.

\section{Discussion}~
We proposed a general scheme which allows one to construct a many-body state which is out of equilibrium on a level of local observables at the two moment of times: the initial $t=0$ and the revival one $t=\tau$. 

The key assumption which we made in the paper is that the unitary which drives the system is similar to a random rotation in a many-body Hilbert space. In particular we assumed that absolute values of $|u_{mn}|\sim 1/\sqrt{2^L}$ are all of the same order and have random phases which are uncorrelated. We believe that this assumption holds for any Hamiltonian which satisfies ETH. We don't have a rigorous proof of this assumption. Nevertheless if we assume that it is correct, then we can suggest a practically useful criteria of ACR reachability. Let us introduce the matrix participation ratio $\text{MPR}$ as 

\begin{align}
\text{MPR}=\left(\sum\limits_{ij}|u_{ij}|^2\right)^2\Biggm/\sum\limits_{ij}|u_{ij}|^4,\nonumber
\end{align}

if for given unitary $u$ the value of $\text{MPR}\rightarrow 1$, then the ACR is reachable. 

The unitary $u$ is not necessarily should describe quantum evolution or be associated with any physical Hamiltonian. The interesting direction is to study ACR for more experimentally relevant unitaries. It is particularly interesting to look at unitaries which can be implemented on existing quantum computers. 

\begin{acknowledgments}
The work is supported by Basis Foundation (Grant No. 18-1-5-19-1).
\end{acknowledgments}


\end{document}